\documentclass[titlepage,twocolumn]{revtex4-1}

\usepackage{bbold}
\usepackage{amssymb}
\usepackage{graphics}
\usepackage{graphicx}
\usepackage{epstopdf}
\usepackage{multirow}
\usepackage{amsmath}
\usepackage{amsthm}
\usepackage{amstext}
\usepackage{amscd}
\usepackage{epsfig}
\usepackage[mathscr]{eucal}
\usepackage{times}
\usepackage{srcltx}
\usepackage{shadow}
\usepackage{url}
\usepackage{color}
\usepackage[all]{xy}

\usepackage[most]{tcolorbox}

\usepackage{hyperref}

\usepackage[english]{babel}

\usepackage[letterpaper,top=2cm,bottom=2cm,left=3cm,right=3cm,marginparwidth=1.75cm]{geometry}

\usepackage{amsmath}
\usepackage{graphicx}
\usepackage{comment}
\begin{document}
\begin{titlepage}
\begin{center}
\line(1,0){420}\\
    \huge{Extinction and Extirpation Conditions in Coalescent and Ecotonal Metacommunities}
\line(1,0){300}\\
\end{center}
\begin{center}
    Martin Heidelman, University of Notre Dame, Department of Physics, mheidelm@nd.edu \\
    Dervis Can Vural, University of Notre Dame, Department of Physics, dvural@nd.edu\\
    \line(1,0){300}\\
\end{center}
    \textbf{Running title:} On Coalescent and Ecotonal Metacommunities
    \break
    \textbf{Statement of authorship:} DCV conceptualized idea. DCV and MH derived analytical results. MH conducted numerical simulations. MH and DCV prepared figures and wrote manuscript. \\
    \break
    \textbf{Data accessibility statement:} Data sharing not applicable – no new data generated as the article describes entirely theoretical research\\
    \break
    \textbf{Article type:} Letter \\
    \textbf{Abstract word count:} 150 \\
    \textbf{Main text word count:} 4968 \\
    \textbf{Reference count:} 47 \\
    \textbf{Number of figures:} 6 \\
    \break
    \textbf{Correspondence to:} Dervis Can Vural, 384G Nieuwland Science Hall, Notre Dame, IN 46556. +1 574-631-6977, dvural@nd.edu)

\end{titlepage}
\null\newpage
\null\newpage

\title{Extinction and extirpation conditions in coalescent and ecotonal metacommunities}
\author{Martin Heidelman, Dervis Can Vural}
\email[]{dvural@nd.edu}
\affiliation{Department of Physics, University of Notre Dame, USA}
\date{\today}

\begin{abstract}
Here we present extinction, extirpation and coexistence conditions where / when two communities combine. We consider one specific model where two communities coalesce, and another model where the communities coexist side by side, blending in a transitionary zone called the ecotone. Specifically, (1) we analytically calculate the shifts in abundances as a function of mixing strength. (2) Obtain a critical value for the mixing strength leading to extinction. (3) Derive an inequality condition for full coexistent mixing. (4) find how the individual communities penetrate into one other as a function of mixing strength. (5) derive the conditions for one species to cross the ecotone and invade an neighboring community and (6) conditions for a native species to get extirpated. Lastly, (7) we spatially investigate the species richness within the ecotone and derive a condition that determines whether the ecotone will have higher or lower richness compared to its surrounding habitats.
\end{abstract}

\maketitle

\textbf{Introduction.}
What happens when two distinct communities fully blend together? Will some species go extinct, or will all species stably coexist? What if two communities came into contact at an interface? In the mixing zone, will the species richness be larger or smaller? How deep will one community penetrate the other? For two such neighboring communities, when can a species from one community jump across and invade the other? Under what conditions will a native species go extinct due to an influx of invaders? 

\begin{figure}[h!]
\centering
\includegraphics[width=0.99\columnwidth]{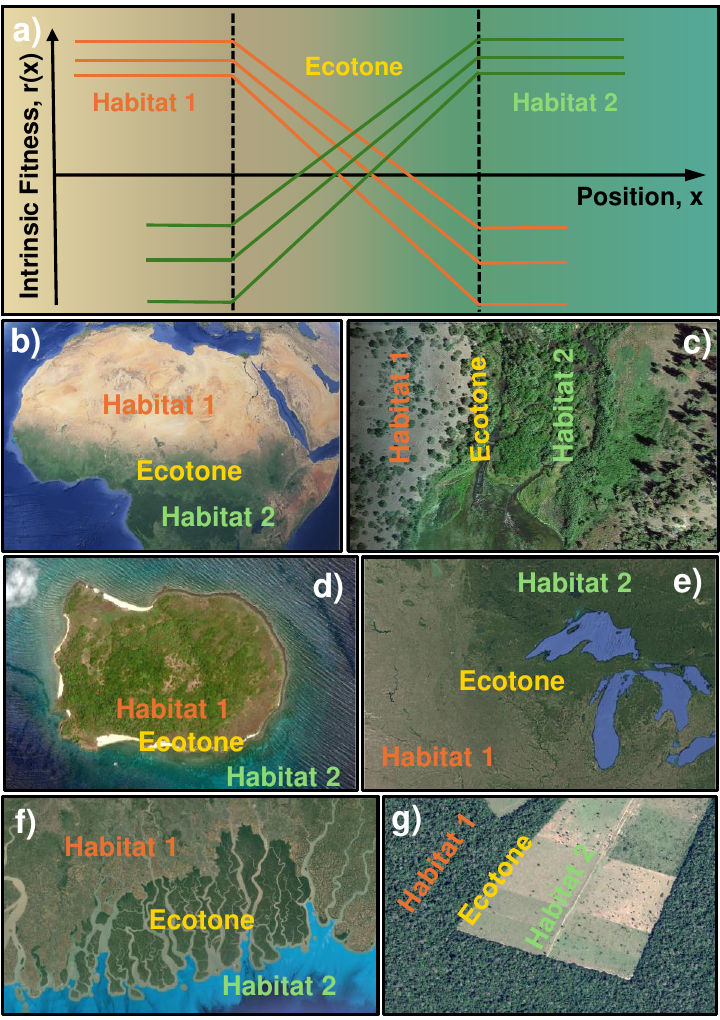}
\caption{\textbf{Ecotones: simplifed model vs. reality} a) The schematics of our modeling approach. For each species, intrinsic fitness $r(x)$ interpolates between high and low values between their native and alien habitats. b) At the biome scale, Africa transitions from desert to rainforest. c) Riparian zones along rivers and streams constitute ecotones. Willow Creek - Missouri River confluence in Montana, shows a transition from the upland ecosystem to the aquatic ecosystem. d) A transition from the tropical upland ecosystem into the deep sea. e) A transition from shrub/grassland ecosystem of the American Midwest into northern Boreal forest. f) The mangrove forests of the Sundarban National Forest in southern India and Bangladesh, a classical example of a coastal ecotone between inland forests and the sea. g) Ecotones are also created anthropogenically, such as this agricultural clearing. The contrast between the full-sun, high disturbance field ecosystem and the low-light forest ecosystem creates a thin boundary where unique community interactions occur. Also visible here is the variation of vegetation cover within different sections of the fields. Each boundary within the field constitutes a micro-ecotone where unique interactions between insects, microbes, fungi, plants etc. that can be described via our framework. Images b-g captured via Google Earth.}
\label{fig:figureZero}
\end{figure}

Community mixing occurs at a multitude of spatial scales \cite{gosz1993ecotone, kolasa1995notes, fortin2000issues, hufkens2009ecotones, hansen2012landscape}. The transitional zone near the natural boundaries between two ecosystems contains a mixture of species from both communities, and is called an ``ecotone'' \cite{levin2009princeton, kolasa1995notes, hufkens2009ecotones, hansen2012landscape}. At the scale of biomes, we see community mixing due to the breakdown of physical barriers from tectonic shifts and uplifts, and climate change induced sea rise \cite{vermeij1991biotas}. Straits, mountain ranges and rivers, or anthropogenic structures such as highways, canals, mountain passes or tunnels can separate two habitats, with a narrow mixing zone in between. An agricultural fence will delineate a forest from a field with very different environmental parameters, thereby creating a narrow ecotonal zone. At microscopic scale, porous media or semi-permeable membranes can create similar transitional zones between microbial communities. Even local variations in light, salinity, pH, temperature and resources can be considered as ecotones. Fig.\ref{fig:figureZero} b-g shows examples to such transitionary zones.

A closely related phenomenon is community coalescence \cite{rillig2015interchange, custer2024toward}. In contrast to an ecotone which delineates two distinct habitats, when communities coalesce, they lose their distinct characteristics and become one well-mixed system. Certain periodic events, such as the seasonal freezing and melting of ice, rising and falling of rivers, or tidal changes, as well as hydraulic infrastructures such as dams, dykes, and weirs, can fully merge two communities. In some cases, the merger can lead to intercommunity interactions that are as strong as intracommunity ones. In other cases where communities meet intermittently through a gap or barrier, the frequency of contact defines the effective interaction strength between the sub-communities. Construction projects merge entire soil microbiomes if soil is moved from one location to another. Ship trade routinely exchanges large volumes of water via ballast discharge.  Ingesting a fermented food, handshaking, kissing, or merely touching a surface in public transport are significant microbiotic merger events. Despite the interesting environmental, agricultural, and biomedical consequences of such mergers, we have only just begun to understand metacommunity dynamics, where one must think beyond species-species interactions and consider community-community interactions \cite{custer2024toward, leibold2004metacommunity}.

While much of metacommunity theory has focused on species dispersal \cite{thompson2017dispersal, fournier2017integrative} and habitat heterogeneity \cite{padmanabha2024landscape} as mechanisms for maintaining species richness, the role of inter-community interactions has also been of interest. Increased interaction strength was shown to decrease dominant species abundance, and an intermediate interaction strength was shown to provide maximal community stability\cite{guichard2005interaction}. Further, it was shown that community stability is influenced by an asynchrony in species interactions between adjacent communities \cite{quevreux2023spatial}. The central parameter of interest here is also the inter-community interaction strength.

Specifically, we investigate the equilibrium properties of coalescent and ecotonal communities as a function of interaction strength / mixing rate / contact rate, and present 7 key results: We analytically describe, and verify with simulations, (1) How abundances shift upon coalescence, (2) What mixing rate leads to extinctions (3) The conditions required for two communities to coexist upon coalescence, and (4) How deep neighboring communities penetrate into one other within the ecotone.

The role of ecotones in the structure and function of the broader landscape has inspired studies for over a century, particularly in regards to conservation efforts, ecosystem services, and biodiversity maintenance. Because ecotones often occur in regions of sharp environmental gradients, they are important zones for speciation \cite{schneider1999test, schilthuizen2000ecotone, freedman2023evidence, kark1999conservation, kark2002peak, kark2007role, smith1997role} 

As such, we (5) Derive the conditions necessary for species crossovers from one habitat to the next, an important processes associated with speciation. We also (6) Study the related phenomenon of extirpation, i.e. the extinction of native species due to an influx of invading species.

A number of ecological characteristics of ecotones have been conjectured, e.g. that they contain a sharp change in vegetation type and/or physiognomy \cite{gosz1992ecological, tansley1953british, walker2003properties}, they feature spatial mosaicity \cite{gosz1992ecological, neilson1992regional, pound1900phytogeography, walker2003properties}, they harbor exotic (ecotonal) species \cite{neilson1992regional,hansen2012landscape, walker2003properties}, and that their species richness can be higher (due to the presence of species from either of the two adjacent communities, either through species movement or through the ``spatial mass-effect'' \cite{leopold1987game, petts1990role, kunin1998biodiversity} or lower (due to being at the limits of a species' range) \cite{van1966relation, van1976establishment, potts2016edge} than on either side of the ecological boundary. However, field data suggests that these patterns are not defining characteristics of ecotones, but emerge as a result of the location's unique species-environment and species-species interactions\cite{walker2003properties, laudenslayer1976breeding, baker2002edge, terborgh1990structure}.

Given such conflicting empirical patterns and theoretical ideas, mechanistic models such as the present work can offer insights into the ecotone species richness puzzle. Accordingly, (7) The species richness within the ecotone as a function of the ecotone parameters will be one of the questions addressed here.

Past methods for modeling inter-community interactions within a metacommunity include the incorporation of a block-structured interaction matrix into classical Lotka-Volterra community dynamics \cite{clenet2023impact, gilpin1994community}. Using this method, \cite{gilpin1994community} found that two communities with a history of competitive exclusion results in a higher probability of an asymmetrical community equilibrium. That is, one community will dominate another, while randomized communities result in a more uniform equilibrium. More recently, \cite{clenet2023impact} used random matrix methods to study the impact of a block structured interaction matrix on the stability, feasibility, and persistence of species. They derived formulas to describe the statistical properties of the extant species, finding that the distribution of abundances followed a truncated Gaussian form. In addition they found that a decrease in community interaction increased the probability of co-feasibility of members of both communities. 

However, the changes in community composition due to small perturbations in inter-community interactions remains unknown. Additionally, an explicit spatial component has been missing from the theory of interacting modular communities. Ecotone and coalescence models that allow for changes in the inter-community interaction strength (say, due to changing environmental conditions, contact area, or contact frequency) would contribute a mechanistic understanding of community composition shifts, which we establish in the present study.

\textbf{Materials and Methods.}
Community dynamics are be modeled by the classical Lotka-Volterra-Fischer equations
\begin{align}
\frac{\partial n_i}{\partial t}=\nabla^2 n_i+n_i\left[r_i(x)+\sum_j A_{ij}n_j\right]\label{eq0}
\end{align}
where $n_i(t,\vec{x})$ and $r_i(\vec{x},t)$ are the abundance and intrinsic fitness of species $i$ at location $\vec{x}$ and time $t$, whereas the matrix $A_{ij}$ quantifies the interaction strength (biomass conversion rate) between species $i$ and $j$. The biotic and abiotic conditions required for these equations to hold true has been well understood \cite{whence}.

When $\vec{r}$ is spatially uniform (or has a linear form that leads to the first term vanishing), the coexistent equilibrium is given by 
\begin{align}
\vec{n}=-A^{-1}\vec{r}\label{equilibrium}
\end{align}

Now, consider two communities with $N_1$ and $N_2$ species, originally out of contact. We can still regard the isolated communities as one system if we define a larger interaction matrix $A$ that contains individual sub-matrices $A_1$ and $A_2$, and an equilibrium abundance vector that contains the abundances of the individual communities
\begin{align}
A=
\left[
\begin{array}{c|c}
A_1 & 0 \\
\hline
0 & A_2
\end{array}
\right] \qquad \vec{n}=\left[\begin{array}{c}\vec{n}_1\\\vec{n}_2\end{array}\right]
\end{align}
where the individual coexistent equilibrium abundances $\vec{n}_1=-A^{-1}_1\vec{r}_1$ and $\vec{n}_2=-A^{-1}_2\vec{r}_2$ are still of size $N_1$ and $N_2$. Now suppose that these two communities come in physical contact. In this case, the total interaction matrix will be modified,
\begin{align}
&A'=A+S=
\left[
\begin{array}{c|c}
A_1 & 0 \\
\hline
0 & A_2
\end{array}
\right]+ \left[
\begin{array}{c|c}
0 & \epsilon S_1 \nonumber\\
\hline
\epsilon S_2 & 0
\end{array}
\right]
\\&\mbox{and abundances shifted, \,}\vec{n}'=\left[\begin{array}{c}\vec{n}_1\\\vec{n}_2\end{array}\right]+\left[\begin{array}{c}\delta\vec{n}_1\\ \delta\vec{n}_2\end{array}\right]\nonumber
\end{align}
where $S_1$ is of size $N_1\!\!\times\!\!N_2$ and describes how the second community influences the first, whereas $S_2$ is vice versa. The parameter $\epsilon$ quantifies the contact frequency, mixing strength, collision rate or the mass conversion rate between two sub-communities.
\begin{figure}[t!]
\centering
\includegraphics[width=\columnwidth]{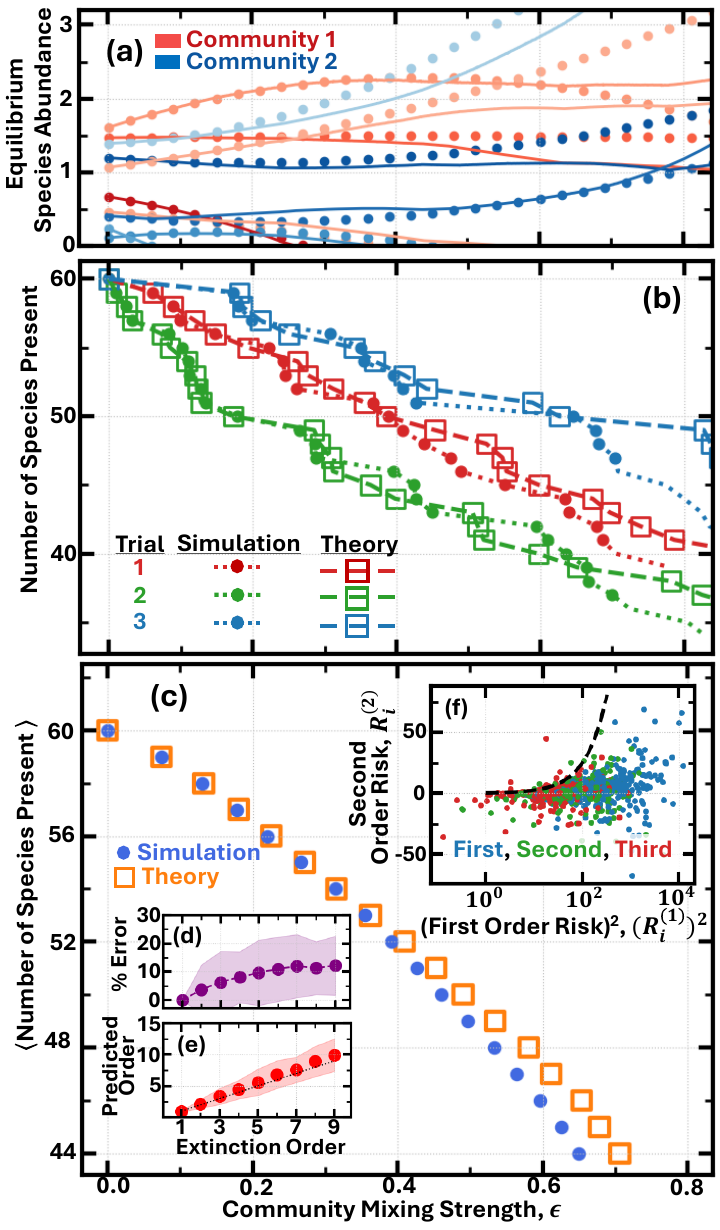}
\caption{\textbf{Community Coalescence.} (a) Equilibrium abundances vs. community mixing strength. Red and blue represent community 1 and 2. Circles show our analytical formulas (up to second order) while the lines show simulations. (b) Surviving species as a function of $\epsilon$ for three representative simulations. Dotted lines and circles show our analytical formulas (up to second order) while dashed lines and open squares show simulations. (c) Number of surviving species vs epsilon averaged over 350 trials. Open squares show averaged analytical results and closed circles show averaged simulation results. (d) Percent error between simulations and analytical formulas for the predicted extinction order. The shaded region shows one standard deviation from the mean. Through the first 5 extinctions we maintain a mean percent error below 10 percent. (e) Mean analytically predicted extirpation order vs. actual (simulated) extirpation order. The dotted line shows the ideal agreement, and the shaded region depicts one standard deviation from the mean. (f) The dashed line delineates inequality Eqn.\ref{condition} for no species loss, $R^{(2)}_i = 4(R^{(1)}_i)^2$. As we see, all first extinctions (blue) are below the dashed line, validating our result. Strictly speaking, Eqn.\ref{condition} is not applicable to ensuing extinctions (green), but we find that it still works quite successfully. No first, 1.4\% of second, and 3.4\% of third extinctions violate the inequality\ref{condition}.}
\label{fig:wellMixed}
\end{figure}

While $\epsilon=0$ would describe two communities completely isolated from each other, $\epsilon=1$ would describe communities that interact just as strongly with each other, as they interact within; and intermediate values, $0<\epsilon<1$ would describe a semi-permeable boundary, a prohibitive distance, or a barrier that is intermittently present. 

As an interesting spatial application, $\epsilon$ can quantify habitat heterogeneity within an ecotone. Ecotones often exhibit mosaicity, so a low $\epsilon$ would describe a mixing region with a spatial arrangement of larger microsites allowing species to mostly interact with their native community members. As such, varying $\epsilon$ can provide some insights into the impact of habitat homogenization. As an interesting temporal application, consider the skin microbiomes of two individuals intermittently in contact. A range of values of $\epsilon$ from 0 to 1 then, would describe no contact, social hand shakes, hand holding, and full contact (perhaps the left and right hand of the same individual). As a second interesting temporal example, the aquatic and terrestrial communities of a riparian ecosystem interact strongly in periods of flooding and interact less under dry conditions. These periodic fluctuations in species interactions will result in periodic changes abundances. For such temporal examples, we consider periodic $\epsilon(t)$'s further below. We will run full population dynamics simulations where two communities are coupled periodically, on and off, as in the case of skin contact or flooding.

\textbf{Community Coalescence.} Before we move on to a spatial model of ecotones, we start with coalescing communities, and calculate how the original abundances $\vec{n}_1$ and $\vec{n}_2$ shift by $\delta \vec{n}_1$ and $\delta \vec{n}_2$ upon mixing.

According to Eqn.\ref{equilibrium} we must invert the combined interaction matrix. \[A^{'-1}\vec{r}=(A + \epsilon S)^{-1}\vec{r}=(I+\epsilon A^{-1}S)^{-1}A^{-1}\vec{r},\] and thus, \[\vec{n}'\!\!=\! (I+\epsilon A^{-1}S)^{-1}\vec{n}=
\!\!\left[
\begin{array}{c|c}
I & \epsilon M_1 \\
\hline
\epsilon M_2 & I
\end{array}
\right]^{-1}\!\!\begin{bmatrix}
    \vec{n}_1    \\
    \vec{n}_2     
\end{bmatrix}
\nonumber
\]
where $M_1 \equiv A_1^{-1}S_1$ and $M_2 \equiv A_2^{-1}S_2$. We now move the matrix to the left and multiply out the sub-matrices to get two equations, $\vec{n}_1' + \epsilon M_1\vec{n}_2' = \vec{n}_1$, and $\vec{n}_2' + \epsilon M_2\vec{n}_1' = \vec{n}_2$, which can be solved:
\begin{align}
\begin{bmatrix}
    \vec{n}_1'    \\
    \vec{n}_2'     
\end{bmatrix}
= 
\begin{bmatrix}
    (I-\epsilon^2 M_1M_2)^{-1}(\vec{n}_1  -\epsilon M_1 \vec{n}_2)     \\
    (I-\epsilon^2 M_2 M_1)^{-1}(\vec{n}_2 -\epsilon M_2\vec{n}_1)      
\end{bmatrix}\label{exact}
\end{align}
As we see, the interactions modify the original abundances by an additive factor determined solely by the abundances of the other community; then a second multiplicative factor mixes these modified abundances together. Eqn.\ref{exact} is exact. It holds regardless of the inter-community interaction strength. However, if the inter-community interactions were weaker than intra-community ones (more precisely, if the eigenvalues of $\epsilon M_{1,2}$ are within the unit circle), we can go further by expanding
\begin{align}
\begin{bmatrix}
    \vec{n}_1'    \\
    \vec{n}_2'     
\end{bmatrix}
= 
\begin{bmatrix}
    \vec{n}_1 - \epsilon M_1\vec{n}_2 +\epsilon^2M_1M_2\vec{n}_1 -   ...   \\
    \vec{n}_2 - \epsilon M_2\vec{n}_1 +\epsilon^2M_2M_1\vec{n}_2 - ...      
\end{bmatrix}.\label{expansion}
\end{align}
which agrees well (Fig.\ref{fig:wellMixed}a) with time dependent simulations of Eqn.\ref{eq0}. This equation also shows us (Fig.\ref{fig:wellMixed}b,c) that as community mixing strength increases, the beta diversity decreases, consistent with empirical data \cite{macarthur1961bird, palmer2003spatial, padmanabha2024landscape}. 

Note that when the interaction between the communities are small compared to those within, the shift in the abundances in one community solely depends on the direct influence of the other, $\delta\vec{n}_1\simeq -\epsilon M_1\vec{n}_2$. For larger interactions, the second order correction should be taken into account, and can be viewed in two parts: the influence of the first community on the second one ($\epsilon M_2\vec{n}_1$), which then hits back the first community (times $\epsilon M_1$). Likewise, for higher order corrections, each time, the $n^{\mbox{th}}$ correction to community-2 must be hit by $\epsilon M_1$ to give the $(n\!+\!1)^{\mbox{th}}$ correction to community 1, and vice versa. In the sense of ``the enemy of a friend of an enemy'', this pattern of successive corrections, each time smaller, push the abundances to their convergent value.

Interestingly, if the interaction between communities approaches a critical value such that one of the eigenvalues of $\epsilon^2 M_1 M_2$ or $\epsilon^2 M_2 M_1$ approaches the unit circle (e.g. when the influence of an enemies enemy is larger than that of an enemy directly), then the successive corrections grow with every correction, and the series diverges and consequently some abundances diverge, while others crash.

We have dealt with similar singularities in earlier works aimed at predicting extinctions in evolving communities \cite{nguyen2021extinction} and developing various community control strategies \cite{nguyen2022theoretical}. As it turns out, such divergences do not constitute a problem when the full time-dependent population dynamics Eqn.\ref{eq0} is solved, however they illegitimize the use of the coexistent equilibrium condition with which we started our analysis. To deal with singular matrices of this kind (in the context here, describing situations where intercommunity interactions are stronger than intracommunity ones), one can remove the rows and columns corresponding to the extinct species until the singularity is removed, and once again move on with the theoretical analysis presented here. 

Fig.\ref{fig:wellMixed}a shows the agreement of our analytical results for the species abundances and critical $\epsilon$'s, with full, time dependent simulations of Eqn.\ref{eq0}.

\textbf{Extinction and coexistence conditions.} The next interesting question to ask is, how strongly should two communities interact before some species go extinct. e.g. how frequently must we shake hands before we lose a species in our microbiome?

To answer this, we set the left hand side of Eqn.\ref{expansion} to zero and solve for $\epsilon$. For community-1, to first order
\begin{align}
\epsilon_{i} = \frac{n_{1i}}{(M_1\vec{n}_2)_i}\label{extinct}
\end{align}
The most endangered species $i$ is the one that would vanish with the weakest contact, i.e. the $i$ for which $\epsilon_i$ is the smallest. This outcome requires a combination of a small numerator and a large denominator. Eqn.\ref{extinct} tells us, very naturally, that the most endangered species are those that have a small abundance to begin with (the numerator), and those that would be most adversely affected by the alien species (the denominator). Note that, since $n_{1i}>0$, species $i$ will have an valid $\epsilon_i>0$ only if the second community has a negative effect on $i$, namely, if $(M_1\vec{n}_2)_i>0$.

As such, we could call the influence per abundance $R_i^{(1)}\equiv(M_1\vec{n}_2)_i/\vec{n}_{1i}$, ``the first order risk''. The second order risk would be $R_i^{(2)}\equiv(M_1M_2\vec{n}_1)_i/\vec{n}_{1i}$, and similarly for higher order terms.

From Eqn.\ref{extinct}, it might appear that any species with with positive first order risk can go extinct upon a sufficiently strong coupling with an alien community. This is not the case. To see why, we set Eqn.\ref{expansion} to zero, and this time solve up to second order, 
\begin{align} 
\epsilon_{i} = \frac{(M_1\vec{n}_2)_i \pm \sqrt{(M_1\vec{n}_2)^2_i - 4 (M_1M_2\vec{n}_1)_i \vec{n}_{1,i}}}{2(M_1M_2\vec{n}_1)_i}\nonumber
\end{align}
which agrees well with simulations (Fig.\ref{fig:wellMixed}b,c); but more interestingly, there are no real solutions when the following condition is satisfied
\begin{align}
(R^{(1)}_i)^2<4R^{(2)}_i\label{condition}
\end{align}
 
This is the condition for no-extinction upon coalescence; and if the equation holds true for all $i$, we expect full coexistence, regardless of how large $\epsilon$. Conversely, the first species to go extinct with increasing $\epsilon$ must always violate this inequality. We have ran a large number of simulations and have observed that the inequality also predicts consequent extinctions reasonably well (inset of Fig.\ref{fig:wellMixed}c).

\begin{figure}
\centering
\includegraphics[width=\columnwidth]{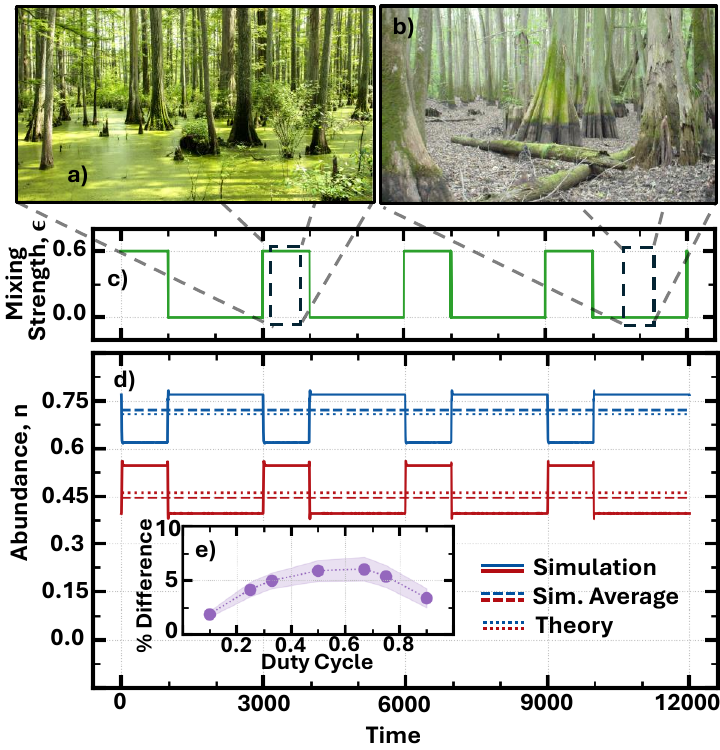}
\caption{\textbf{Intermittent Community Mixing.} Communities may interact at various strengths depending on whether a riparian forest is (a) flooded or (b) dry. We run Lotka-Volterra simulations where $\epsilon(t)$ varies periodically between 0 and 0.6 and show that the time averaged abundances can be described with our analytical formulas using an effective mixing strength intermediate in value. (c) $\epsilon(t)$ for a duty cycle of $1/3$ (d) Abundances of two species $n_i(t)$. Solid lines show the fluctuating abundances according to simulations, dashed lines show the time average that of, and the dotted lines show the prediction of our analytical formula. (e) The small difference between analytical and simulation results as a function of duty cycle length. Photo Credit: a) Ralph Earlandson, b) Hans-Christian Rohr}
\label{fig:epsilonVtime}
\end{figure}

\textbf{Intermittent Community Mixing.}
As discussed earlier, communities may come into contact intermittently, causing time dependent shifts in abundances. To accurately describe such cases, we define a time averaged mixing strength
\begin{align}
\epsilon=\epsilon_1 \tau_1/T+\epsilon_2\tau_2/T, \label{effective}
\end{align}
where $\tau_{1,2}$ are high and low mixing durations, $\epsilon_{1,2}$, are the high and low mixing strengths, and $T=\tau_1+\tau_2$ is the cycle period. Plugging this effective $\epsilon$ into our analytical formulas, we find good agreement with time dependent simulations of Eqn.\ref{eq0}, where $\epsilon(t)$ is set to a square wave (Fig.\ref{fig:epsilonVtime}).

\textbf{Ecotones.}
So far, we studied how two communities merge into one. However, they can also stand side by side, blend at the boundary, but otherwise remain spatially distinct. Terrestrial and aquatic species for example, confine to their respective niches, but they can interact along the coastline. To describe such spatially heterogeneous systems, we must introduce a location-dependent intrinsic fitness, $r(x)$.

Specifically, we define three regions of space: the ``left'' habitat, the ecotone, and the ``right habitat''. We assume that species have a higher intrinsic fitness in their native habitat, and a lower one in the alien habitat --to the degree that they cannot survive in the latter. In the ecotone region in between, we extrapolate between these high and low values. The form of $r(x)$ we adopt is shown in Fig.\ref{fig:figureZero}a, and based on this assumption, we get to write the (zero $\epsilon$, unperturbed) abundances as
\begin{align}\begin{split}
\vec{n}_1(x) = \vec{n}_{1L} + \frac{1}{L}(\vec{n}_{1L}-\vec{n}_{1R})x, \\
\vec{n}_2(x) =  \vec{n}_{2L}+ \frac{1}{L}(\vec{n}_{2R}-\vec{n}_{2L})x. \label{n(x)}
\end{split}
\end{align}
Here, $x=0$ is where the left habitat ends and the ecotone begins, and $L$ is the ecotone thickness, beyond which extends the right habitat. The subscripts $L,R$ mark the left/right habitat locations and the indices $1,2$ mark the community of species native to the these left and right habitats. For example the components of $\vec{n}_{1R}$ denote the abundances of the first community on the right (their non-native) habitat.

These abundances indeed satisfy our assumptions. First, a linear $\vec{r}(x)$ will give a linear $n(x)$ because the diffusion term in Eqn.\ref{eq0} will vanish and $\vec{n}(x)=-A^{-1}\vec{r}(x)$, and then the linearity of $\vec{r}(x)$ implies the linearity of $\vec{n}(x)$. Secondly, we get the correct (zero $\epsilon$, unperturbed) abundances at the boundaries $\vec{n}_{1L}=-A_1^{-1}\vec{r}_{1L}$, $\vec{n}_{1R}=-A_2^{-1}\vec{r}_{2R}$, $\vec{n}_{2L}=-A_2^{-1}\vec{r}_{2L}$ $\vec{n}_{2R}=-A_1^{-1}\vec{r}_{1R}$, since there are no alien species in a habitat in the absence of interactions.

As trivial as Eqns.\ref{n(x)} may seem, they give rise to problems if taken literally: These linear forms dip below zero towards alien habitats, indicating that the edges of the ecotone are so inhospitable that they can only support a ``negative abundance'' of a certain species. These superfluous negative abundances will interact with the actual species, incorrectly altering their abundances (e.g., rabbits will benefit from the negative sharks on land!). Note that Eqn.\ref{eq0} itself does not suffer from this problem, since the $n_i$ multiplying the square bracket curbs the decay of abundances, preventing them from ever falling  below zero. The negative abundance problem simply arises because we insist upon imposing the coexistent equilibrium condition Eqn.\ref{equilibrium} where there is no coexistence (say, of rabbits and sharks, at the shore).

\begin{figure*}[t!]
\centering
\includegraphics[width=\linewidth]{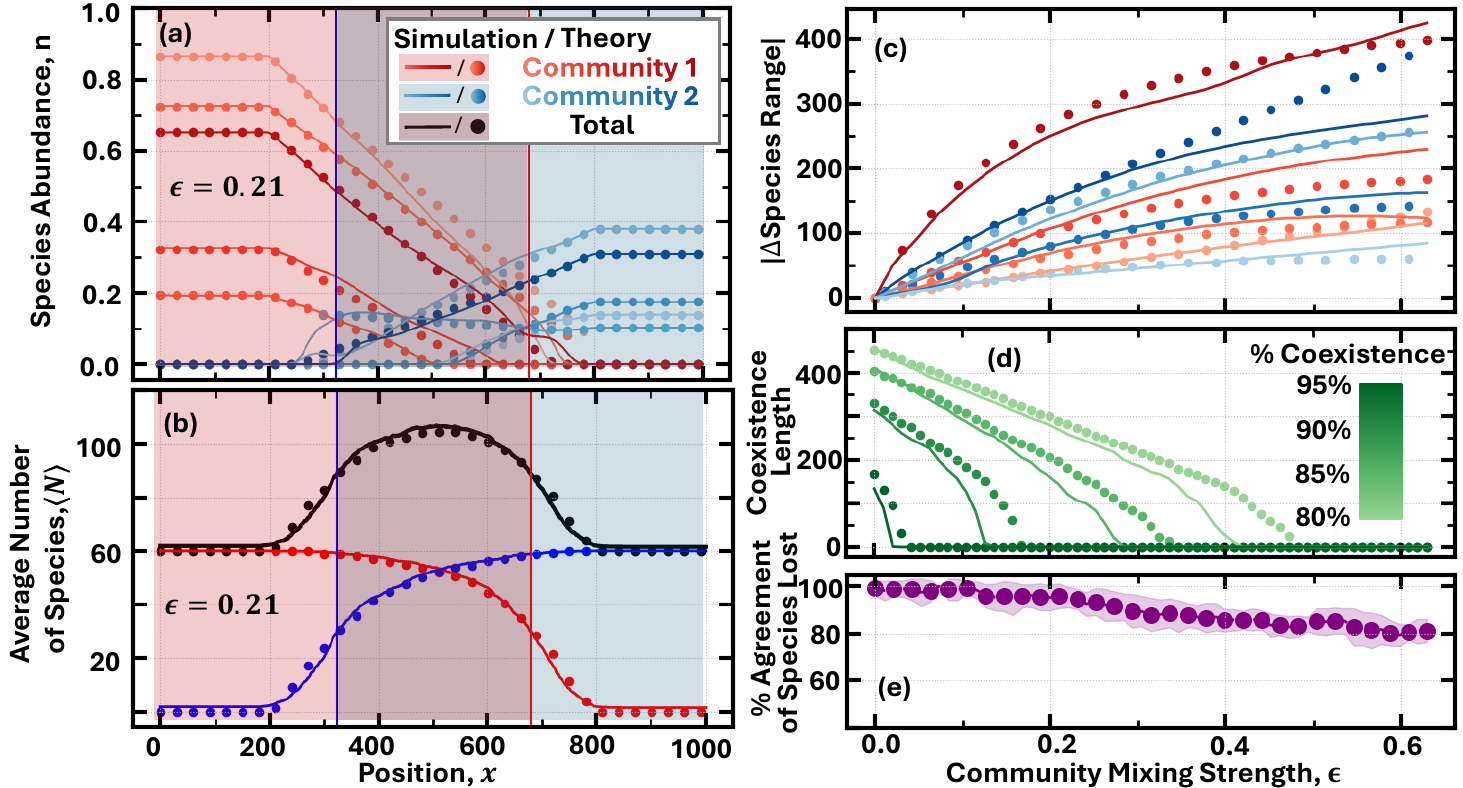}
\caption{ \textbf{Summary of the spatial ecotone model.} a) Species abundance vs. position for a community mixing strength of $\epsilon = 0.21$. The abundances of 10 of 120 total species are shown, with species of community 1 shown in the red palette and species of community 2 shown in the blue palette. Numerically simulated abundances are shown by the solid line, and analytically predicted abundances are shown by the solid circles. For both a) and b), Red and blue shaded regions depict the spatial zones where 50 percent or more of the species from communities 1 and 2 are present respectively. b) Average number of species present vs, position for a community mixing strength of $\epsilon = 0.21$. Red and blue curves are for species of communities 1 and 2 respectively and black curves are for the total species between both communities. Solid curves are for numerical and scatter points are for analytical results. Averages are taken over 30 separate trials.c)Absolute change in species range beyond their native habitat for 10 select species. Curves, symbols, and colors are the same as described in a). d) Coexistence lengths vs. community mixing strength for four levels of coexistence, between 80(light green)-95(dark green) percent coexistence. Solid lines depict simulation results and filled circles are for our theoretical results. e) The average percent agreement between analytical and numerical results of species lost vs. community mixing strength. The shaded region depicts one standard deviation from the mean. Each point represents the average percent of species that our analytical formulas correctly predicts to be lost at the center of the model domain with respect to those predicted by the numerical simulations.}
\label{fig:LinearFixedL}
\end{figure*}

Fortunately, we have a practical (albeit post-hoc) workaround to resolve this issue: We ignore the presence of negative abundances, go ahead and substitute Eqn.\ref{n(x)} into Eqn.\ref{expansion}, get our result, but then, after the fact, eliminate all the influence of these negative abundances from our formula. Up to first order,
\begin{align}
\begin{bmatrix}
    \vec{n}'_1    \\
    \vec{n}'_2     
\end{bmatrix} = 
\begin{bmatrix}
    [\vec{n}_1 - M_1(\vec{n}_2*\Theta(\vec{n}_2))]* \Theta(\vec{n}_1)  \\
    [\vec{n}_2 - M_2(\vec{n}_1*\Theta(\vec{n}_1))]* \Theta(\vec{n}_2)
\end{bmatrix}\label{thetaFunctions}
\end{align}
where $*$ is a component-wise vector multiplication, and $\Theta$, our post-hoc fix, is the Heaviside function, which returns 0 for negative arguments, and 1 for positive ones. On both rows, the leftmost Heaviside functions ensure that no negative abundances have an effect on positive abundances, whereas the rightmost ones set all negative abundances to zero. We see, in Fig.\ref{fig:LinearFixedL}a that this analytical result agrees well with simulations of Eqn.\ref{eq0}.

How much will a species' range expand or retract as a function of the coupling strength between the two communities? How much do two communities penetrate one other as a function of mixing strength? To find out species' ranges, we set $n'(x)=0$ in Eqn.\ref{thetaFunctions}
\begin{align}\label{absoluteLocation1}
 \vec{x}_L(\epsilon)= \frac{\vec{n}_{1L}-\epsilon M_1\vec{n}_{2L}}{\vec{\delta}_1 + \epsilon M_1\vec{\delta}_2}L
\end{align}
which gives us the range expansion with varying $\epsilon$,
\begin{align}\label{range}
\Delta \vec{x}_L(\epsilon)=\frac{\epsilon L [\vec{\delta}_{1}*M_1 \vec{n}_{2L}+M_1\vec{\delta}_2*\vec{n}_{1L}]}{\vec{\delta}_{1}*(\vec{\delta}_{1}+\epsilon M_1\vec{\delta}_2)}
\end{align}
where in Eqn.\ref{absoluteLocation1} and Eqn.\ref{range} $*$ and $/$ are both component-wise operations, and $\vec{\delta}_{1}=\vec{n}_{R1}-\vec{n}_{L1}$ and $\delta_2$ are the difference in abundances between habitats, for community one and two. The shift of range for the right community is given by the same formula, with subscripts $1,2$ and $L,R$ exchanged.

Plugging in Eqn.\ref{absoluteLocation1} below gives us the species richness as a function of position,
\begin{align}
\!\!\!\!\!N(x) \!=\! N \!-\! \sum_{i}\Theta(x\!-\!x_{L,i})+\sum_{i} \Theta(x\!-\!x_{R,i})\label{speciescount}
\end{align}
where $N=N_1+N_2$. In Fig.\ref{fig:LinearFixedL}b, we verify this result with simulations and also show the first and second terms separately, which count the left and right species as a function of position (red and blue curves).

In Fig.\ref{fig:LinearFixedL}c, we see how the range of species change as a function of mixing strength, and in Fig.\ref{fig:LinearFixedL}d, we use these range equations to calculate a ``coexistence length'', that is, the distance in which more than a certain fraction of species coexist. In Fig.\ref{fig:LinearFixedL}d we plot the coexistence length versus mixing strength for four coexistence thresholds between $80\%$ and $95\%$. In Fig.\ref{fig:LinearFixedL}e, we show the percent agreement between the species lost according to simulations and versus our analytical formulas.

\textbf{Species crossover.} Favorable interactions can allow a species access to an otherwise hostile alien habitat. Therefore, as the interaction / mixing strength between two communities increases, we expect to see some species expand their range across the ecotone, and ultimately, even invade the alien habitat.

In the absence of interactions, there are no species of community-1 in the right habitat, $\vec{n}_1(x)<0$ for $x>L$. As we turn up $\epsilon$, the equilibrium abundances will start changing, and at a critical $\epsilon$ value, one specific invasive species, which we label by $k$, will have its equilibrium abundance $n_k$ rise above zero in the alien habitat. At this critical point, the equilibrium relation reads
\begin{align}\label{crossoverMatrix}
\left[
\begin{array}{c|c}
A_{1,kk} & \epsilon S_{1k} \\
\hline
\epsilon S_{2k} & A_2
\end{array}
\right]
\begin{bmatrix}
    n_k'    \\
    \vec{n}'_{2R}     
\end{bmatrix}=-
\begin{bmatrix}
    (\vec{r}_{1R})_k   \\
    \vec{r}_{2R}     
\end{bmatrix}
\end{align}
where $n_k$  and $A_{1,kk}$ are the abundance and self-interaction of the crossover species, $A_2$ is the full $N_2 \times N_2$ interaction matrix of community 2, and $S_{1k}$ and $S_{2k}$ are the $k^{\mbox{th}}$ row and $k^{\mbox{th}}$ column of $S_1$ and $S_2$ respectively. Note that here, $\vec{n}_{2R}$, is the vector of native species abundances prior to community mixing. 

What is the abundance of $k$? We multiply out the first row of the matrix, substitute $n_{1R,k}=r_{1R}/A_{1,kk}$ and $M_1= A_1^{-1}S_1$, and write it all in vector form,
\begin{align}\label{crossover abundances}
\vec{n}'(\epsilon)=\vec{n}_{1R}-\epsilon M_1 \vec{n}_{2R}
\end{align}
The components of the left hand side vector are the abundances of the potential invaders.

For small $\epsilon$ values, the equilibrium abundance of our potential invader is negative in the alien habitat, as expected. But once $\epsilon$ reaches a critical value, one species $k$ in Eqn.\ref{crossover abundances} will stop being negative and hit zero. This will happen for the species $k$ for which 
\begin{align}\label{ecrit}
\epsilon_{\mbox{crit,}k} = \frac{(\vec{r}_{1R})_k}{\sum_{j}S_{1,kj}(\vec{n}_{2R})_j}
\end{align}
is the smallest. Here, we wrote $(\vec{n}_{2R})_j$ without a prime because at $\epsilon_{\mbox{crit,}k}$ no alien species has a positive abundance to affect the native community. To obtain analogous expressions for the crossover abundances and critical $\epsilon$ for the other community, we must simply replace subscripts 1 with 2, and L with R.

Perhaps unsurprisingly, we have recovered the first order result of the well-mixed model in Eqn.\ref{crossover abundances}, because we only took into account the interaction between invaders and the native community, and neglected the interactions between invaders.

\begin{figure}[t!]
\centering
\includegraphics[width=\columnwidth]{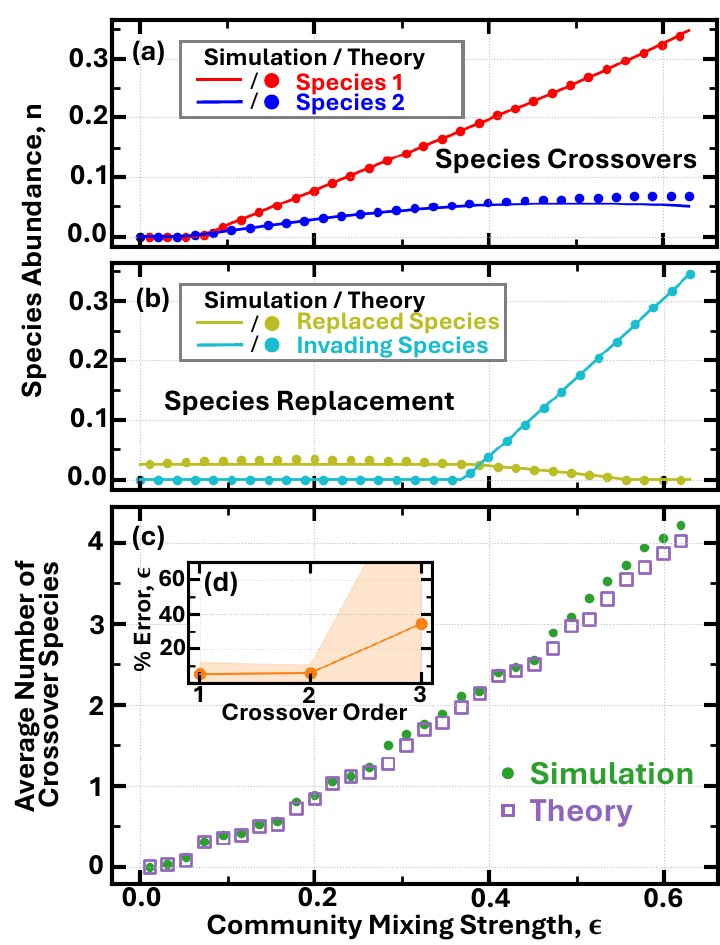}
\caption{\textbf{Species crossover and replacement.} (a) Species abundance of the first species to crossover into the alien community as a function of community mixing strength, $\epsilon$. (b) Species abundance as a function of $\epsilon$ for a species replacement event. For both panels (a) and (b), scatter points show results from our analytical formulas and the curves show results from numerical simulation. (c) Average number of crossed over species versus community mixing strength, $\epsilon$. Green scatter points show our analytical averages and open purple squares depict numerical averages. Averages are over 36 trials. (d) percent error in the critical $\epsilon$ for species crossover as predicted between our analytical formulas and the numerical results versus the number of species crossed into the alien habitat versus the crossover order. Even though our derivation for species crossover was only intended to predict the abundance of the first species to crossover, we predict up to the second species crossover with an average of 5 percent error. Crossovers beyond the second are beyond our prediction ability.}
\label{fig:crossovers}
\end{figure}

What about multiple invaders for even larger $\epsilon$ values? In this scenario, we could still argue that the $k$ for which $\epsilon_{\mbox{crit},k}$
is the second largest, will crossover second, and so on. However, we must be careful that using this formula for multiple invaders will neglect the interactions between the invaders. This will be fine when the number and/or abundance of the invaders are few. Eqn.\ref{crossover abundances} and Eqn.\ref{ecrit} agrees well with numerical Lotka-Volterra simulations in describing the first couple of crossovers (Fig.\ref{fig:crossovers}).

\textbf{Extirpations.} Upon increasing the inter-community interactions further, some species may go extinct in their native habitats due to an influx of invaders. This phenomenon is called extirpation. There are important prerequisites for extirpation in our model. First, because all species in their native habitat begin at equilibrium abundances, for a species to become extirpated, at least one alien must crossover, thereby altering the prior stable equilibrium. Second, because the order of species crossover is not known a priori, one must first calculate the critical epsilons for crossover, order these epsilons, and then include only these terms in the derivation. While this is possible to accomplish numerically, the analytical utility diminishes beyond the first crossover. For this reason we restrict our derivation of extirpation formulas to include only a single invader, replacing a single native species. 

To do so, we plug in Eqn.\ref{crossover abundances} into Eqn.\ref{crossoverMatrix}, and designate one of the native species $l$ of the second community to get extirpated with increasing $\epsilon$. This yields, up to first order,
\begin{align} \label{replacement}
(\vec{n}'_{2R})_l = (\vec{n}_{2R})_l +\epsilon \frac{M_{2,kl}(r_{1,R})_k}{A_{1,kk}}.
\end{align}

And since $(r_{1,R})_k/A_{1,kk} = -(\vec{n}_{1,R})_k$, we get $\vec{n}'_{2,R} = \vec{n}_{2}-\epsilon(M_1 \vec{n}_{1,R})$ as was the case in the well-mixed community coalescence model.
The critical $\epsilon$ can then be found by setting $(\vec{n}'_{2,R})_l =0 $ and solving for $\epsilon$ in Eqn.\ref{replacement},
\begin{align} \label{Criticalreplacement}
\epsilon_{\mbox{crit}} = \frac{(\vec{n}_{2,R})_l}{M_{2,kl}(\vec{n}_{1,R})_k}.
\end{align}

We find good agreement between this equation and Lotka-Volterra simulations (Fig.\ref{fig:crossovers}).

\textbf{The effect of ecotone parameters on species richness.} \label{Vary_L}
We return to one of our original questions, of whether an ecotone has more or less species richness compared to its surroundings. In our framework, this entirely depends on the parameters describing the spatial dimension $L$ and the fitness profiles $\vec{r}(x)$. We consider here two cases of varying ecotonal parameters. The first, Case-A, corresponds to a situation where the fitness values on each side is kept constant while the habitats themselves are displaced. Imagine for example, two aquariums, each with different temperature, salinity, pH and light levels, connected by a tube, which constitutes the ecotone. Case-A amounts to keeping the intrinsic fitness values at the boundaries $\vec{r}_{1,2}(0)$ and $\vec{r}_{1,2}(L)$ fixed, while varying $L$ so that the slope of $r(x)$ changes in between. The consequences of this variation are rather trivial: Since we would be simply stretching the entire picture horizontally by some stretch factor $a$, all instances of $x$ (in $r(x)$ and our results for $n(x)$ and $\epsilon_{\mbox{crit}}$) must simply be replaced, $x\to a x$. 

\begin{figure*}[t!]
\centering
\includegraphics[width=\linewidth]{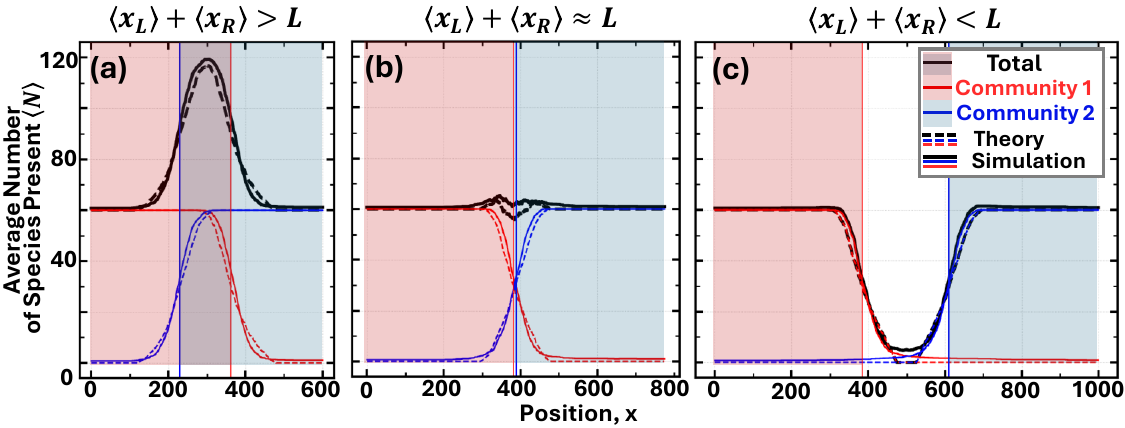}
\caption{Average number of species present vs position for three different mixing lengths. For all, the number of species in the left and right community were fixed at 60 for a total of 120 species. Averages were taken across 12 data sets with a fixed mixing strength of $\epsilon = 0.2$. Vertical red/blue lines show the position where community 1 and 2 species number drops to half of the initial value respectively. Red/blue shaded regions depict where at least 50 percent of right/left community species exist. Dashed lines depict second order formulas, solid lines represent numerical results. (a) Mixing length of 360. Here, the spatially ``forcing'' increases species diversity in the mixing region. (b) Mixing length of 535. Here, spatial ``forcing'' is just right to maintain species diversity across the model. (c) Mixing length of 760. Here, the communities are too far apart to increase the species diversity between them.}
\label{fig:ChangeLL}
\end{figure*}

A second, less trivial scenario, Case-B, is where the conditions in two communities differentiate from each other such that the fitness of the species decrease in the alien habitat while they remain constant in their native home. This might occur due to shifts in climate at evolutionary timescales, where native species adapt to their new environmental conditions, while their fitness in the alien habitat lessens further. In Fig.\ref{fig:ChangeLL}, we utilize Eqn.\ref{speciescount} to determine the average number of species versus spatial position for this Case-B. We keep $\vec{r}_1(x=0)$ and $\vec{r}_2(x=L)$ constant while varying $L$, so that $\vec{r}_1(x=L)$ and $\vec{r}_2(x=0)$ shifts lower. As we see, the species richness across the mixing region has a strong dependence on the length of the mixing region, (or the degree of dissimilarity between the two ecosystems), and numerical simulations agree well with our analytic results. 

We are now prepared to give a condition for species richness maintenance within the ecotone. If the distance at which the left and right community looses half its species is longer than the size of the ecotone itself, then the tails of left and right species number function overlap, and the  species richness would be larger or equal everywhere compared to had there been only one habitat. In other words, if
\begin{align}
\langle x_L\rangle+\langle x_R\rangle\geq L\label{biodiversity}
\end{align}
where the angle brackets denote median extirpation length of the left and right species, as determined by Eqn.\ref{absoluteLocation1}. The working principle behind the inequality is illustrated in Fig.\ref{fig:ChangeLL}.

We find that, for a constant mixing strength parameter, an increase in species diversity (relative to the number of species present in either adjacent community) occurs when the two environments are more similar, and are close in proximity.

Conversely, we predict a decline in species diversity the more dissimilar the two adjacent environments are to one another. This prediction, however, needs to be taken with an understanding that our models do not include the introduction of new species, or the time adaption of species traits. For example, after site disturbance, the number of species in the mixing region between two drastically different adjacent ecosystems, perhaps in the riparian zone between aquatic and terrestrial ecosystems may be low initially after site disturbance. However, over time, fitter species may begin to colonize the newly cleared zone, and the diversity of species in the mixing region should be expected to depend on the type and the diversity of propagule inputs, weighted by the ability of each species to colonize the unoccupied land. This lends to the historical debate of the importance in making the distinction between an ``ecotone'' and an ``ecocline'',  where more heavily disturbed sites, such as riparian zones subjected to periodic flooding events, are predicted to hold relatively low levels of species diversity as compared with more stable ecotones such as forest-grassland, or prairie-desert ecotones \cite{van1990ecotones, van1976establishment, van1966relation}.

\textbf{Discussion.} 
We calculated how strongly two communities must mix before species loss occurs Fig.(\ref{extinct}) and conversely, established an inequality for safe community coalescence (Eqn.\ref{condition}). We have shown the utility of our formulas for intermittently coupling communities in addition to static ones (Eqn.\ref{effective}). We have given the change in species range in an ecotone as a function of mixing strength (Eqn.\ref{range}), and the species richness of the ecotone in terms of these ranges (Eqn.\ref{speciescount}). 

One of the more interesting predictions of our model has been that of species crossovers  (Eqn.\ref{ecrit}) and extirpations (Eqn.\ref{replacement}) above a critical mixing strength. The crossover phenomenon exemplifies the speciation capacity of ecotonal regions. High diversity of landscape features within a small spatial proximity allows a species to sample a variety of habitats each with its own unique set of species interactions. As the strength of these interactions changes, either through evolutionary or environmental processes, then the species can migrate from one region into another. With evolutionary time, a species can then undergo allopatric speciation, a drastic example of which, is the trading of fins for limbs during the fish-tetrapod transition of vertebrates in the Devonian geologic period \cite{schwarz2023using, buatois2022invasion}, and more recently, the speciation of an African rainforest skink \cite{freedman2023evidence}. 

Lastly, we have derived an inequality for an ecotone to have higher species richness than either of its neighboring habitats standing alone (Eqn.\ref{biodiversity}). Interpreting the mixing strength parameter as a measure of habitat heterogeneity, our results are consistent with previous observations and theory of increased species diversity with increasing habitat diversity \cite{macarthur1961bird, palmer2003spatial, padmanabha2024landscape}, but see \cite{cramer2005habitat}.

\textbf{Simulation Details.} While our analytical results work for all interaction matrices and fitness parameters, simulations require choosing a large number of parameter values and certain numerical methods Eqn.\ref{eq0}. For the coalescence model, the system of equations was integrated using the solve\_ivp function of the scipy.integrate library using RK45, an explicit Runge-Kutta method of order 5. For the spatial models, a custom class is written using the py-pde open source python package to integrate the equations, using the method of lines by explicitly discretizing space using a finite difference scheme, and the time evolution is carried out using a simple Euler scheme. A one-dimensional Cartesian grid was initialized from $0$ to $L$ with $Nx$ spatial support points. Boundary conditions were set to fixed value for species in their native community, and zero derivative in the alien community. Initial conditions sent to the solver were the first order analytical results. The solution time, $T=1200$, was set to an arbitrarily large value where convergence was ensured and held consistent across trials. The py-pde adaptive time-stepper was used.

For all simulations, each interaction matrix element was chosen from a uniform distribution between $-0.5$ and $0.5$, were made anti-symmetric by averaging each matrix with its transpose, and then positive elements were reduced by an inefficiency factor of $0.65$. Diagonal elements of the intracommunity matrices were set to $-1$. For spatial simulations, a diffusion constant of $1$ was used for all species.

For the data shown in Fig.\ref{fig:epsilonVtime}, communities were equal in size ($N_1 = N_2 = 30$). Initial species abundances were chosen from a uniform random distribution between $0.1$ and $1.0$. $\epsilon$ ranged from $0$ to $0.6$. For intermittent mixing simulations, seven different $\epsilon(t)$ duty cycles were simulated, between $0.1$ and $0.9$. Panel \ref{fig:epsilonVtime}c), depicts averages over 100 separate trials at each duty cycle. 

In figure \ref{fig:wellMixed}, communities were equal in size ($N_1 = N_2 = 30$), and initial abundances were chosen from a uniform random distribution between $0.1$ and $2.0$. The averages in panel \ref{fig:wellMixed}c)-e), were taken over 350 separate trials.

In Fig.\ref{fig:LinearFixedL}, \ref{fig:crossovers}, and Fig.\ref{fig:ChangeLL}, $N_1 = N_2 = 60$, and Native abundances were chosen from a uniform random distribution between $0.01$ and $1.0$, and their abundances in the alien community were chosen from a uniform distribution from $-0.1$ to $-0.001$.

\bibliographystyle{alpha}
\bibliography{main}
\end{document}